\documentclass[9pt]{article}
\usepackage[a4paper, bottom=1cm]{geometry}
\usepackage{endnotes}
\usepackage{graphicx}
\usepackage{fullpage}
\usepackage{setspace}
\usepackage{lineno}
\usepackage{url}
\usepackage{subfigure}
\usepackage{epstopdf}
\usepackage{cite}
\usepackage{graphicx}
\usepackage{mdwtab}
\usepackage{mdwmath}
\usepackage{multicol}
\usepackage{amssymb, amsmath}
\usepackage{color}
\usepackage{txfonts}
\usepackage{CJK}

\begin{document}
\title{What are the limits on Commercial Bank Lending?}
\author{Jacky Mallett (jmallett@ieee.org}

\maketitle
\section*{Abstract}
Analysis of the 2007-8 credit crisis has concentrated on issues of relaxed lending
standards, and the perception of irrational behaviour by speculative investors in
real estate and other assets. Asset backed securities have been extensively criticised
for creating a moral hazard in loan issuance and an associated increase in default risk, by removing
the immediate lender's incentive to ensure that the underlying loans could be repaid.
However significant monetary issues can accompany any form of increased commercial bank lending, 
and these appear
to have been overlooked by this analysis. In this paper we propose a general explanation for 
credit crises based on an examination of the mechanics of the banking system, and
in particular its internal controls on the supply of credit.
We suggest that the current credit crisis is the result of multiple failures in the Basel
regulatory framework, including the removal of central bank reserve requirements from some
classes of deposit accounts within the banking system, allowing
financial instruments representing debt to be used as regulatory capital, and in particular
the introduction of securitized lending which effectively removed a previously implicit 
control over the total quantity of
lending originating from the banking system.  We further argue that the interaction of these 
problems has 
led to a destabilising imbalance between total money and loan supply growth, in that
total lending sourced from the commercial bank sector increased    
at a faster rate than accompanying growth in the money supply. This not only created a multi-decade 
macro-economic debt spiral, but by increasing the ratio of debt to money within the monetary
system acted to increase the risk of loan defaults, and consequentially reduce the overall
stability of the banking system.
\section*{Introduction}
Research on the causes of the current series of multi-national credit crises has implicated securitized 
lending, and other innovative financial instruments, primarily for removing the burden of the financial 
consequences of loan default from the issuers of successfully securitized loans as discussed by Diamond\cite{diamond.2009}, 
and for the creation of an opaque system of interlinked 
borrowing as described by Gorton\cite{gorton.2008}. Nyberg also cites general systemic problems in the banking 
system\cite{nyberg.2011}, and evidence presented by Schularick and Taylor\cite{schularick.2011} also
points to rapid credit expansions being a strong predictor of financial crises since 1870.
There has however 
been relatively little focus on the precise reason why securitized lending originating within the 
banking system, and the creation of accompanying trading markets in securitized debt should cause such 
large scale systemic problems, whereas a long history of corporate and government
bond debt instruments in a similar trading environment has not been considered similarly problematic.
\par
\begin{figure}[h]
\begin{center}
\includegraphics[width=10cm]{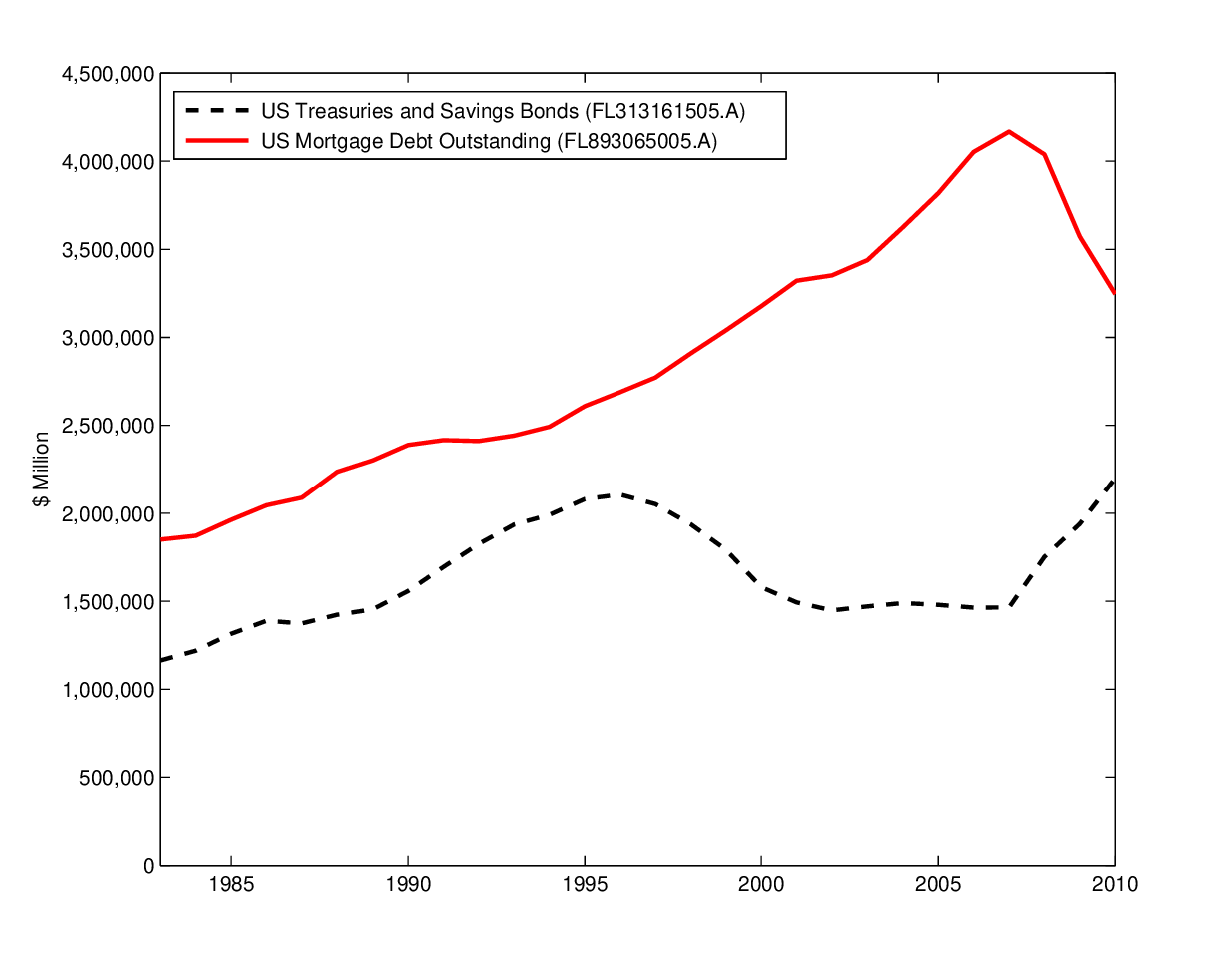}
\caption{US Mortgage Debt vs US Government Debt}
\label{fig_usdebt}
\end{center}
\end{figure}

Although public attention is often drawn to seemingly dramatic increases in government debt, a larger
source of borrowing growth in many countries appears to have been debt originating from the banking system.
While part of the increase in both forms of debt is due to the underlying growth of the money supply, 
it is clear from 
Figure \ref{fig_usdebt}\footnote{Data from the Federal Reserve Bank`s Flow of Funds Z.1 series. 
Care has to be exercised with all monetary data owing to the continuous expansion of the unit of 
measurement, i.e. the money supply.
Data in this figure has consequently been
normalised for growth in the money supply with a base of 1983, using the M2 money
supply measure (Table H.6) with the Money Market funds component removed. (Money market funds are primarily
held in short term debt instruments. Total US money supply growth for this period was approximately 
4 times.)} 
which shows mortgage debt for the United States over the last 25 years, in comparison with the total amount
of outstanding US Government Treasuries, that mortgage debt has been increasing at a significantly 
faster rate.  Growth in banking debt 
considerably exceeded that of government borrowing up until
the credit crisis of 2007, at which point a sizeable increase in government debt occurred as a result of
intervention to rescue the American banking system from collapse.
\par
The question of why bank lending was able to expand so considerably, and by
extension what limits apply to it, is of particular interest in understanding
the periodic excesses in lending and borrowing that have come to be termed credit
bubbles. If there is no limit on lending originating from the banking system, and
the growth in credit is being supplied from there, then there should be no particular 
mystery about the cause of credit bubbles. Any increase in the supply 
of debt is liable to trigger a simple positive feedback loop if the asset being
purchased is in limited supply. If housing developers are
borrowing money to buy land for example, then the price they can pay is determined
by the supply of loans. If an increase in bank lending causes more money to be created 
to purchase these assets, then the price of the asset will rise, amplifying the signal 
that asset prices provide to investors, and increasing the demand for loans. 
The resulting feedback loop is likely to be particularly vicious if the asset 
can then be used as collateral for another loan, either through sale, or remortgaging against
the new value of the property, since
a direct debt/price feedback loop will be created. The 1920's stock market
bubble in the US provides an interesting example of this where margin loans were used to finance 
share purchases, with the shares so purchased then acting as collateral for further borrowing.
\par
Conversely if there is a limit on the total quantity of bank lending, but expansion of the
loan supply from the commercial banks has not reached it, then a credit bubble could
be regarded as an unusually large variation in credit demand within known limits.
In this case policy towards controlling the damage caused by credit bubbles could be simply
directed to detecting and preventing this increase. It would then be straightforward
to detect asset bubbles simply by monitoring changes in the supply of credit to the economy.
\par
A possible explanation for credit bubbles then is that current theoretical models of banking
equilibrium are incorrect, and that in practice failures in banking
regulation or the invention of novel financial instruments and banking practices allow the system  
to enter periods of unrestricted loan supply growth, with major and consequent distortions in that 
part of the money and loan supply that is controlled by institutions performing reserve
based lending. To explore this explanation in the context of the 2007 credit crisis, 
we will first discuss
the theoretical model of the banking system and its deficiencies, and the current situation
with respect to the Basel regulatory frameworks.
We will then
explore the specific features of some financial instruments, notably Asset Backed or Mortgage
Backed securities, which challenge the integrity of this framework, and examine
the empirical evidence for these effects.
\section{Theoretical Models of the Banking System}
\subsection{Fractional Reserve Based Lending}
Banking as we understand it today has emerged over several centuries from a set of practices first 
established in Northern Europe by medieval goldsmiths and traders\cite{quinn.1994}. It initially 
developed as a form of statistical multiplexing whereby access to physical money in the form of 
gold was managed through day to day bookkeeping practices, operated under the assumption that 
only a fraction of the underlying liabilities (customer deposits of gold) would be requested at any one time. 
Based on this assumption goldsmiths would make short term loans to other customers, and as the 
chits used to represent gold deposits began to be exchanged directly, a bank based monetary system developed. 
Over time this system has mutated into today`s almost entirely electronic transfer based system, 
while still retaining the bookkeeping practices of the original system. 
\par
The standard description of this system found in most textbooks
describes a simple recursive process where loans are made against deposits,
and the money they represent is re-deposited into the banking system, 
creating a damped recursive feedback process 
that acts to expand the total quantity of bank loans and bank deposits.
As each new deposit is created a required fraction of the
deposit is held back as a reserve by the bank, with the remaining deposits providing backing
for further loans.  Since its inception this relationship between money and lending within the banking
system has consequently exerted an independent influence on the price level to that
performed by physical money whenever transfers are made from within the banking
system, for example by writing cheques\footnote{Since Gold Standard era regulation rested
on prescribed relationships between physical notes and coins, and the \emph{price} of gold reserves,  
the influence of direct transfers within the banking system on the general price level would have
been a destabilising factor.}
\par
\begin{figure}[h]
\begin{center}
\includegraphics[width=7cm]{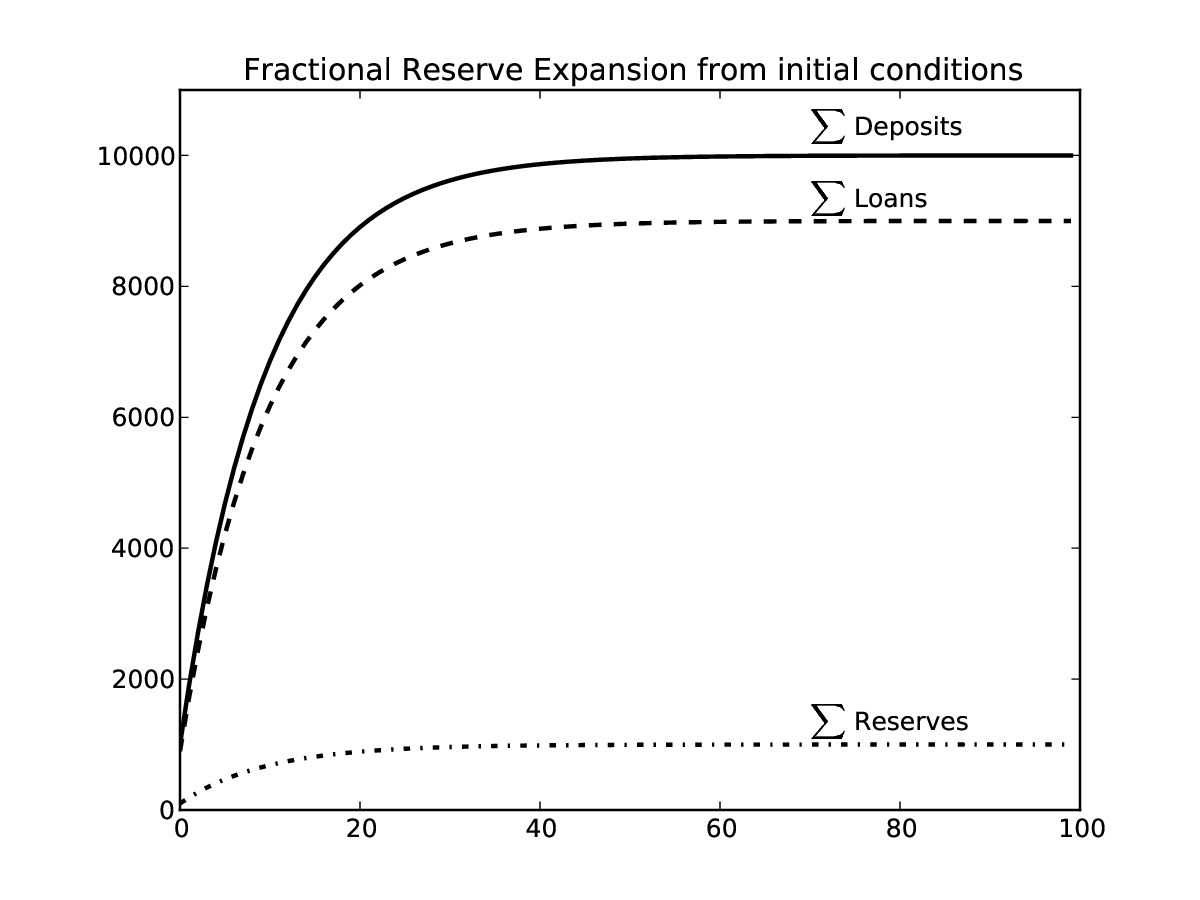}
\vspace*{8pt}
\caption{Theoretical Expansion of the Money Supply from initial conditions.}
\label{fig_frprocess}
\end{center}
\end{figure}
\par
\begin{figure}[h]
\begin{center}
\includegraphics[width=7cm]{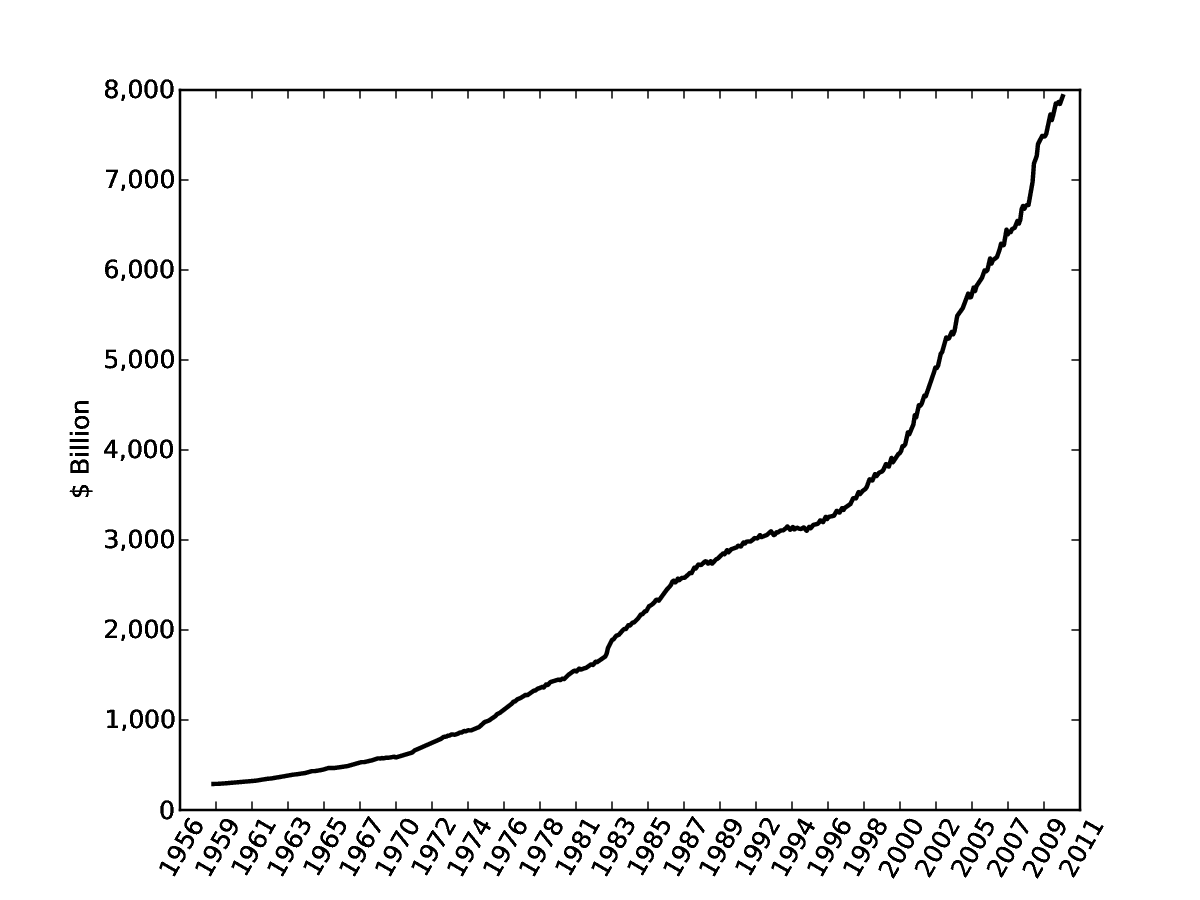}
\vspace*{8pt}
\caption{US M2 with Money Market funds removed: 1959 - 2010}
\label{fig_h6hist5}
\end{center}
\end{figure}

Figure \ref{fig_frprocess} shows the theoretical behaviour of the banking system as described in 
foundational economic textbooks such as Mankiw\cite{mankiw.1997}. 
With a reserve requirement of 10\% the eventual expansion of 
the money supply is 10 times the original deposit into the system, and that of the loan 
supply, 9 times the original deposit. Under this theoretical model, 
the bank originated loan supply
is always constrained to be a fraction of the money supply, 
90\% in the example shown, and cannot exceed the money supply. The actual behaviour 
of the US money supply over the last 40 years is shown in
Figure \ref{fig_h6hist5} and of the loan supply in Figure \ref{fig_absusassets}.
\par
The simplified textbook description appears to have originated in the
1931 Macmillan report to the British Parliament\cite{macmillan.1931} and
appears to have been authored by Keynes\cite{stamp.1931}.
It did not include either loan repayments or loan defaults, and was probably
only intended as a simplified explanation for the deposit expansion process the implications of which
were being explored by economists at that time\cite{lawrence.1928}. It incorrectly suggests
that reserves were a fraction withheld from the customer's deposits, in reality the fractional reserve
is an additional amount of money held by banks on deposit at the central bank. 
\par
One correct implication of the textbook model 
is that the limit on commercial bank loans would always be a 
fraction of the total amount of deposits in the banking system, and that
bank lending would expand or contract in a fixed proportion with
the quantity of deposits.  In the textbook model, once expansion reached    
the limits imposed by reserve requirements, new loans could only be
made as the principle on existing loans was repaid, freeing up loan capacity.
Both the money and loan supplies could vary within these limits though. 
For example if loans are repaid faster than they are made 
then the money supply will contract as the expansion process reverses.
This behaviour is rarely seen, but is economically dangerous
since although market based mechanisms can over time adjust most prices in the economy, they do
not apply to fixed monetary contracts such as loans.
Hart\cite{hart.1938} in his analysis of the behaviour of the 
US banking system during the later years of the 1930's Great Depression 
attributes net repayment of debt with removing more money from the money
supply than the earlier wave of bank failures had.\footnote{ Money supply induced deflation 
should not however be confused with 
price deflation resulting from increases in the supply of goods, which is generally a sign of increased 
prosperity as more goods become available at cheaper prices.}
\par
Loans within the economy can consequently be classified into two different types,
transfer loans such as bonds and treasuries, which are a direct 
exchange of debt for money, and reserve based loans which are
issued by banking institutions. 
The distinction between the two types of lending is important,
since operations on the two types of debt are not equivalent.
Some financial operations such as 
transfer or sale have no impact within the larger monetary system when performed on transfer
loans, but can cause systemic side effects when performed on commercial bank loans, 
as will be shown below. 
\subsection{Basel Capital Reserve Based Lending}
In most countries today the operational and regulatory framework of
the banking system has changed significantly from the
theoretical model described above, particularly with respect to
the fractional reserve requirement which has been increasingly removed. 
In the American system the full reserve requirement is limited to 
"Net Transaction Accounts"\footnote{In the USA, reserve requirements are 10\% on
Net Transaction accounts, and 0\% on non-personal time deposits and Euro currency
liabilities. \url{http://www.federalreserve.gov/monetarypolicy/reservereq.htm}}, which
are a small proportion of total deposits. The focus of bank regulation has shifted to the
"Capital Reserve Requirements", which represent a mixture of shareholder equity, reserves against
loss, and subordinated debt the total quantity of which is
based on a complex series of risk based analyses of the bank's loan book. \cite{schooner.2010}.
\begin{figure}[ht]
\begin{center}
\includegraphics[width=12cm]{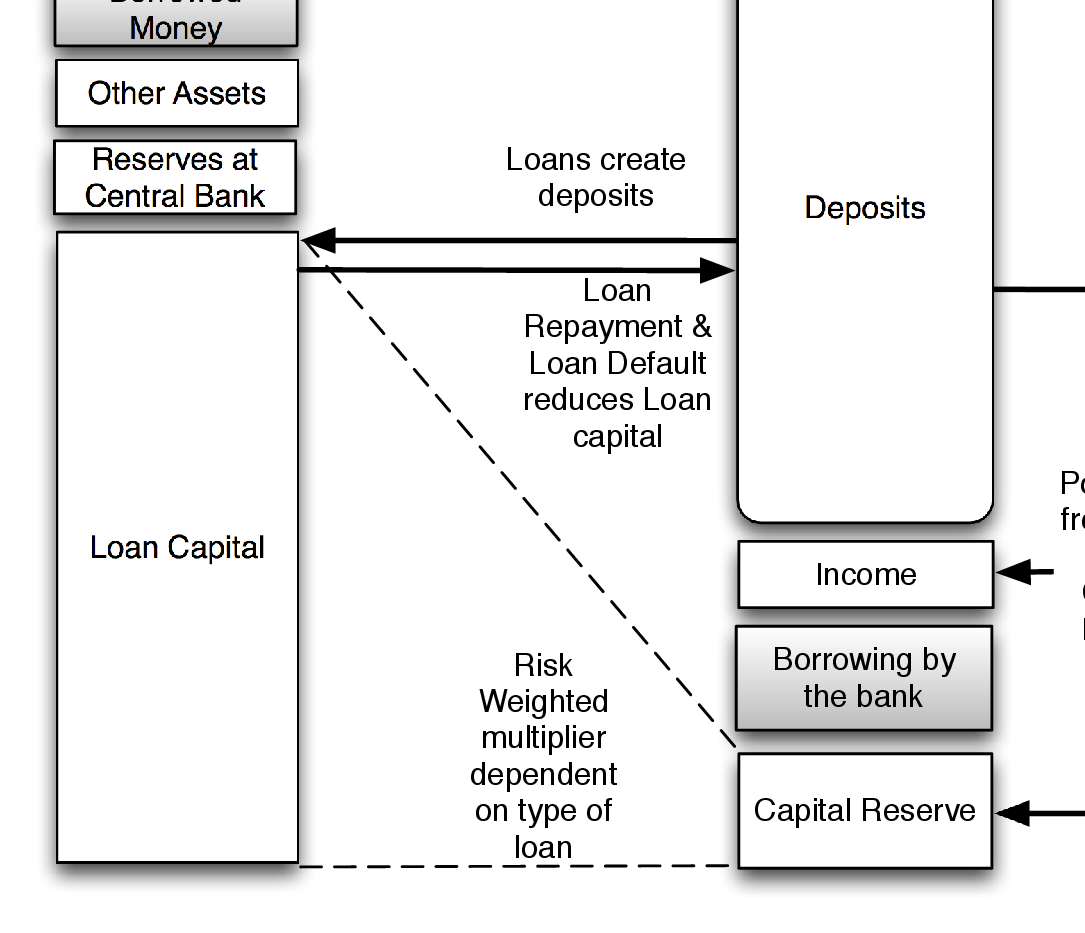}
\vspace*{8pt}
\caption{Regulatory framework of Basel Based Banking Systems.}
\label{fig_basel}
\end{center}
\end{figure}
\par
When losses occur on a bank's loans, the amount lost is first covered from profits
and then from regulatory capital. If these are not sufficient, bank failure and loss of bank deposits
can occur. Consequently the Basel accords concentrate on regulating the 
capital reserve requirements of individual banks on a risk weighted basis in order
to prevent bank failure through inadequate loss provision. 
The full title of the Basel Accords is the "International Convergence
of Capital Measurement and Capital Standards", they do not attempt to 
provide regulation over money creation
through the lending mechanism, nor any form of systemic regulation of the behaviour
of the entire banking system.  
\footnote{Pillar 2 of the Basel Accords is a commitment to regulatory enforcement
and overview. However, this pre-supposes that the behaviour of the underlying system is 
sufficiently well understood that it can be appropriately regulated.}
\par
The accords approach this task by establishing minimum requirements for the ratio of the value of the
capital reserve in relation to the total loan capital of loans made by the bank. 
To be well-capitalised under American federal bank regulatory definitions,
which is a requirement to avoid extra supervision, a bank's regulatory capital holding
must be at least 10\% of its risk adjusted loans.  Risk adjustment applies to the type of the loan
being made, for example mortgages have a risk
adjustment weighting of 50\%, so banks can lend twice as much 
in proportion to their regulatory capital holdings for mortgages than for other types of loans.
As Acharya points out\cite{acharya.2010} this can bias banks towards certain types of loans, 
and in particular 
Basel II assigned government bonds rated AA- or higher a 0\% risk weighting.  
Changes in risk ratings can also be highly problematic owing to the accompanying multiplier
effects on allowed lending.
\par
Figure \ref{fig_basel} shows a simplified illustration of this framework within the context
of the accounting framework maintained at each bank, with arrows indicating some of the 
potential feedback loops.
The total quantity of loans issued by a bank in the diagram is effectively regulated by 
the multiple of lending allowed by its risk weighted capital(shown as dashed lines), and 
the local status of its $liabilities = assets + equity$ accounting identity. 
Loans that have been sold by the banks
to non-bank entities(i.e. MBS and ABS)  are not shown on the diagram.
\par
The expansion of money in the form of bank deposits under this framework is nominally
constrained by the combination of reserve requirements at the central bank, the capital reserve, and
the amount of money in customer accounts on deposit at the bank. Reserve requirements
at the central bank are limited in their effectiveness, both by the presence of deposit
accounts that do not have a reserve requirement, by the ability of banks to borrow from each 
other to support their reserve requirements, and by the necessity of government and central
bank intervention should reserves prove insufficient for the entire system. There are no
limits on the total amount of capital reserves that can be held by the entire banking
system, and capital holdings can be increased from profits as required. Customer deposits
are being continuously created and removed by the act of loan creation and repayment, 
and will increase while the rate of new lending exceeds the rate of loan repayment and default.
Monetary and loan supply contraction is also a possibility if the rate of new lending falls
below this level. Taken together, it would seem that equilibrium behaviour would be an 
unlikely outcome for this system.
\subsection{Modern Monetary Theory}
Attempts within Economics to develop a more complete theory of the banking system
are broadly categorised under "exogenous" theories where deposits create loans, 
and "endogenous" or post-Keynsian theories where lending
creates deposits. Interestingly many endogenous money 
authors such as Lavoie\cite{lavoie.2003} continue the tradition set by Keynes of omitting
loan repayment and default, and there seems to be a general aversion to presenting the historical 
context of a system whose rules vary considerably between time and place.
The focus on credit also frequently overlooks the role of money in establishing the 
price level, and thereby providing a critical economic signal, and the unfortunate reality 
that credit and money are directly linked within the banking system is rarely addressed simultaneously.
Although Minsky's financial instability hypothesis\cite{minsky.1992} does consider the role of
financial debt relationships 
relative to financial activity, it does not consider the operational mechanics of the banking
system in sufficient detail to uncover the intrinsic mechanisms for this instability. 
\par
Both theories of money contain accurate observations of some aspects of the
banking systems' behaviour, but neither offers a complete system analysis, and in a recursively defined 
system to debate whether loans create deposits or deposits create loans is quintessentially tautologous. 
Further problems are created by their inclusion
within economic theories of supply and demand for money and credit, and definitions of money as
debt that can be traced to the origins of money as bills of account. 
This is particularly noticeable with the proponents 
of monetary circuit theory\cite{rochon.2003-1} and 
the idea that there is a demand for money and credit
which determines the system's behaviour. Certainly historically there have been 
periods where physical shortages of money caused significant
problems, and Fischer\cite{fischer.1996} provides a fascinating account of such problems in the medieval 
ages. However one of the considerable advantages of the banking system was that it 
alleviated these issues, and discussion in particular of the demand for money tends to overlook
the interaction with the price level that occurs when the money supply itself increases.
Debates over the demand for credit also fail to consider that the system itself 
provides a very clear signal when insufficient 
demand occurs(it contracts), absent which behaviour it can be safely assumed that limits on loan supply are
dominating.
\par
Even a simple simulation of the textbook model with loan repayments
suggests that this is a system whose behaviour is sensitive to many conditions\cite{mallett.2011}, and
it is probably this failure to consider the multiple and separate causes of gross macro-economic
features such as price deflation and credit expansion originating from 
seemingly minor differences between banking systems that have created significant problems for any
purely empirically based analysis.
Determining how any given currency's banking system will behave over time is challenging,
complete and testable descriptions of their regulatory frameworks are not currently provided
by any central bank.  There is also considerable confusion
in the monetary statistics themselves, with no consistency in either the measures being 
used, or the components used within them. The definition of M2 used for the Euro for example,
is significantly different from that provided by the Federal Reserve Banks for the US dollar, and
differs again from that used in other countries.
\section{Failures in Systemic Regulation}
\subsection{Asset Backed Securities}
Asset Backed Securities(ABS) are freely traded financial instruments 
that represent packages of loans issued by the
commercial banks. The majority are created from mortgages, but credit 
card debt, commercial real estate loans, student
loans, and hedge fund loans are also known to have been securitized\cite{rosenthal.1988}
The earliest form of ABS within the American banking 
system\footnote{Ludwig Von Mises,\cite{mises.1912} writing in 1912 describes the 
Austro-Hungarian Bank "refusing to issue bills ... to increase the
difficulty of speculative repurchase of home securities from abroad",
suggesting that a loan sale problem may also have been present at that time.}
appears
to stem from the creation of the Federal National Mortgage Association
(Fannie Mae) in 1938 as part of amendments to the US National Housing Act, 
a Great Depression measure aimed at creating loan
liquidity. Fannie Mae, and the other 
Government Sponsored Enterprises
buy loans from approved mortgage sellers, typically banks, and create
guaranteed financial debt instruments from them, to be sold on the
credit markets.  The resulting bonds, backed as they are by loan
insurance,  are widely used in pension
funds and insurance companies, as a secure, financial instrument
providing a predictable, low risk return.
\par
The creation of a more general form of Mortgage Backed Security   
is credited to Bob Dall and the trading desk of 
Salmon brothers in 1977 by Lewis\cite{lewis.1990}. 
Lewis also describes
a rapid expansion in their sale beginning in 1981 as a side effect of the
United States savings and loans crisis. The concept was extended in 
1987 by bankers at Drexel Burnham Lambert Inc. to corporate bonds and loans
in the form of Collateralized Debt Obligations(CDOs), which
eventually came to include mortgage backed securities, and
in the form of CDO-Squared instruments, pools of CDO. (For an excellent
work on the history of these instruments see Mackenzie's paper
on the sociology of the knowledge of these instruments within the
financial community\cite{mackenzie.2011}.)
\par
Analysis of the systemic effects of Asset Backed Security has   
concentrated chiefly on their ability to improve
the quantity of loans, or loan liquidity, which has been treated as a 
positive feature by Greenspan\cite{greenspan.2005}.  Wolfe\cite{wolfe.2000} also noted that
securitization allowed banks
to increase their return on capital by transforming
their operations into a credit generating pipeline process,
but didn't consider the effects of this mechanism on the larger
banking system. Shin\cite{shin.2009} has also
analysed their effect on bank leverage and the stability of the larger
financial system within 
an accounting framework. He highlights the significance
of loan supply factors in causing the sub-prime crisis. Although 
his model appears not to completely incorporate the full implications of the process
operating within the capital reserve regulated banking system as described below, it presents
an alternate, matrix based analysis of the uncontrolled debt expansion that
these instruments permit.
\par
The systemic problem introduced by asset backed securities, or any form of sale
that transfers loans made by commercial banking institutions outside the regulatory
framework is that they allow banks to escape the explicit reserve and 
regulatory capital based regulation on the total amount of loans being issued against
customer deposits. This has the effect of steadily increasing the ratio of bank originated
loans to money on deposit within the banking system.
\par
The following example demonstrates the
problem using two  banks, A and B. 
For simplicity fees related to loans and ABS sales are excluded. 
It is assumed that the deposit accounts are Net Transaction accounts
under current US banking requirements\footnote{Feinman\cite{feinman.1993} has 
a review of current and previous reserve requirements within the US system.},
and therefore carry a 10\% reserve requirement, and that both banks 
are "well capitalised" under FDIC and 
Basel treaty definitions, and that the risk weighted
multiplier for the capital reserve for these loans is also 10.
\par
The example
proceeds as a series of interactions as money flows between the two banks.
The liabilities (deposits) and assets (loans) are shown, with loans being
separated into bank loans, and Mortgage Backed Securities(MBS), depending
on their state.
\paragraph{Initial Conditions:}
To simplify Bank B is shown as having made no loans, and has excess reserves at the central
bank to maintain the balance sheet.
The normal inter-bank and central bank lending mechanisms would enable the bank to 
compensate for temporary imbalances during the process under normal conditions. 
All deposit money used within the example remains on deposit at either Bank A 
or Bank B. On the right hand side of the table the total amount of deposits and loans for both banks   
is shown.
\par
\begin{table}[h]
\centering
{
\caption{Unregulated Increase in Leverage through Loan Sales}
\label{table_1a}
\begin{tabular}{|lccc|cc||cc|}
\hline
\multicolumn{8}{|c|}{Initial Conditions} \\
\hline
     &\multicolumn{3}{c|}{Assets}                      &\multicolumn{2}{c||}{Liabilities}&     &            \\
Bank & Loans         & MBS                & CB Reserve& Deposits& Capital   & $\sum$ Deposits & $\sum$ Loans\\
\hline
A:   & 1000 	     & 0                  &  100      &  1000   &    100    &  2000           & 1000        \\
B:   &               &                    &  1100     &  1000   &    100    &    	      &             \\
\hline
\end{tabular}
}
\end{table}

%\newpage
\paragraph{Step 1:}Bank A creates a \$1000 Mortgage Backed Security from the loan 
on its balance sheet.          

\begin{table}[!h]
\centering
{
\caption{Unregulated Increase in Leverage through Loan Sales}
\label{table_1b}
\begin{tabular}{|lccc|cc||cc|}
\hline
\multicolumn{8}{|c|}{Step 1} \\
\hline
     &\multicolumn{3}{c|}{Assets}                      &\multicolumn{2}{c||}{Liabilities}&     &            \\
Bank & Loans         & MBS                & CB Reserve& Deposits& Capital   & $\sum$ Deposits & $\sum$ Loans\\
\hline
A:   &               & 1000               &  100      &  1000   &    100    &  2000           & 1000        \\
B:   &               &                    &  1100     &  1000   &    100    &                 &             \\
\hline
\end{tabular}
}
\end{table}
\paragraph{Step 2:}
The securitized loan is sold to the \emph{depositor} at Bank B. 
\$1000 is paid to Bank A from that depositor in payment for the loan. Bank A
now has no loans outstanding against its deposits, and the securitized
loan has been moved outside of banking system regulation. Note that total
deposits at the two banks have temporarily shrunk due to the repayment of
the loan capital at A. The actual transfer of the deposits between the banks is 
facilitated through the reserve holdings which also function as clearing funds.

\begin{table}[h]
\centering
{
\caption{Unregulated Increase in Leverage through Loan Sales}
\label{table_1c}
\begin{tabular}{|lccc|cc||cc|}
\hline
\multicolumn{8}{|c|}{Step 2} \\
\hline
     &\multicolumn{3}{c|}{Assets}                      &\multicolumn{2}{c||}{Liabilities}&     &            \\
Bank & Loans         & MBS                & CB Reserve& Deposits& Capital   & $\sum$ Deposits & $\sum$ Loans\\
\hline
A:   &               &                    &  1100     &  1000   &    100    &  1000           & 1000        \\
B:   &               &                    &  100      &         &    100    &                 &             \\
\hline
\multicolumn{8}{|l|}{\emph{Mortgage Backed Securities: \$1000}} \\
\hline
\end{tabular}
}
\end{table}
\paragraph{Step 3} As Bank A now has no loans against its deposits,
and is within its regulatory capital ratios, it
can make a new \$1000 loan. The funds from 
this loan are deposited at Bank B. The sum of the deposits
rises as a result as does the quantity of loans. Note that the transfer of the loan money
from Bank A to Bank B again goes through the reserve holdings in the clearing system and
restores the original balance at Bank B.

\begin{table}[h]
\centering
{
\caption{Unregulated Increase in Leverage through Loan Sales}
\label{table_1d}
\begin{tabular}{|lccc|cc||cc|}
\hline
\multicolumn{8}{|c|}{Step 3} \\
\hline
     &\multicolumn{3}{c|}{Assets}                      &\multicolumn{2}{c||}{Liabilities}&     &            \\
Bank & Loans         & MBS                & CB Reserve& Deposits& Capital   & $\sum$ Deposits & $\sum$ Loans\\
\hline
A:   & 1000          &                    &  100     &  1000   &    100    &  2000           & 2000        \\
B:   &               &                    &  1100    &  1000   &    100    &                 &             \\
\hline
\multicolumn{8}{|l|}{\emph{Mortgage Backed Securities: \$1000}} \\
\hline
\end{tabular}
}
\end{table}
%\newpage
\paragraph{Step 4}
Bank A securitizes the loan made in Step 3 repeating \emph{Step 1.}
\begin{table}[h]
\centering
{
\caption{Unregulated Increase in Leverage through Loan Sales}
\label{table_1e}
\begin{tabular}{|lccc|cc||cc|}
\hline
\multicolumn{8}{|c|}{Step 4} \\
\hline
     &\multicolumn{3}{c|}{Assets}                      &\multicolumn{2}{c||}{Liabilities}&     &            \\
Bank & Loans         & MBS                & CB Reserve& Deposits& Capital   & $\sum$ Deposits & $\sum$ Loans\\
\hline
A:   &               & 1000               &  100     &  1000   &    100    &  2000           & 2000        \\
B:   &               &                    &  1100    &  1000   &    100    &                 &             \\
\hline
\multicolumn{8}{|l|}{\emph{Mortgage Backed Securities: \$1000}} \\
\hline
\end{tabular}
}
\end{table}
\newpage
\paragraph{Step 5}
Bank A sells the MBS to the \emph{depositor} at Bank B, repeating \emph{Step 2.}
\begin{table}[h]
\centering
{
\caption{Unregulated Increase in Leverage through Loan Sales}
\label{table_1f}
\begin{tabular}{|lccc|cc||cc|}
\hline
\multicolumn{8}{|c|}{Step 5} \\
\hline
     &\multicolumn{3}{c|}{Assets}                      &\multicolumn{2}{c||}{Liabilities}&     &            \\
Bank & Loans         & MBS                & CB Reserve& Deposits& Capital   & $\sum$ Deposits & $\sum$ Loans\\
\hline
A:   &               &                    &  1100    &  1000   &    100    &  1000           & 2000        \\
B:   &               &                    &   100    &         &    100    &                 &             \\
\hline
\multicolumn{8}{|l|}{\emph{Mortgage Backed Securities: \$2000}} \\
\hline
\end{tabular}
}
\end{table}
\paragraph{Step 6}
Bank A makes a new loan which is deposited at Bank B, repeating \emph{Step 3}
\begin{table}[h]
\centering
{
\caption{Unregulated Increase in Leverage through Loan Sales}
\label{table_1g}
\begin{tabular}{|lccc|cc||cc|}
\hline
\multicolumn{8}{|c|}{Step 6} \\
\hline
     &\multicolumn{3}{c|}{Assets}                      &\multicolumn{2}{c||}{Liabilities}&     &            \\
Bank & Loans         & MBS                & CB Reserve& Deposits& Capital   & $\sum$ Deposits & $\sum$ Loans\\
\hline
A:   & 1000          &                    &   100    &  1000   &    100    &  2000           & 3000        \\
B:   &               &                    &  1100    &  1000   &    100    &                 &             \\
\hline
\multicolumn{8}{|l|}{\emph{Mortgage Backed Securities: \$2000}} \\
\hline
\end{tabular}
}
\end{table}
%\newpage
\paragraph{Step 7}
Bank A securitizes the loan made in Step 6, repeating \emph{Step 4}
\begin{table}[h]
\centering
{
\caption{Unregulated Increase in Leverage through Loan Sales}
\label{table_1h}
\begin{tabular}{|lccc|cc||cc|}
\hline
\multicolumn{8}{|c|}{Step 7} \\
\hline
     &\multicolumn{3}{c|}{Assets}                      &\multicolumn{2}{c||}{Liabilities}&     &            \\
Bank & Loans         & MBS                & CB Reserve& Deposits& Capital   & $\sum$ Deposits & $\sum$ Loans\\
\hline
A:   &               & 1000               &   100    &  1000   &    100    &  2000           & 3000        \\
B:   &               &                    &  1100    &  1000   &    100    &                 &             \\
\hline
\multicolumn{8}{|l|}{\emph{Mortgage Backed Securities: \$2000}} \\
\hline
\end{tabular}
}
\end{table}

\begin{center}
\emph{Since the Deposit and Loan positions of the two banks are identical
in all respects in Steps (1,4), (2,5), (3,6) and (4,7) the process 
can continue indefinitely, resulting in expansion of the total commercial
bank originated loan supply independent of central bank control.}
\end{center}
This is a simplified version of the flows between loans, deposits, 
and asset backed securities that occur daily in the banking system.
At no point has either bank needed 
recourse to central bank funds, or broken any of
their statutory requirements with respect to capitalisation or 
reserve requirements where they apply. 
\par
The problem is the implicit assumption with reserve based 
banking systems that bank originated loans remain within the banking system.
Allowing the sale of loans 
to holders outside of the regulated banking system (i.e.
to entities other than regulated banks) removes these loans from that
control and thus creates a systemic 
loophole in the regulation of the commercial bank loan supply.
\par
The introduction of loans sales has consequently created a novel situation 
in those modern economies that allow them, not only in
causing a significant expansion in total lending from the banking sector, but also in changing
the systemic relationship between the quantity of money in the system to the quantity of bank
originated debt, and thereby considerably diluting the influence the central bank can exert
over the loan supply.
The requirement that no individual bank should lend more than 
their deposits has been enforced by required reserves at the central
bank since the 19\textsuperscript{th} century in Europe, and the early 20\textsuperscript{th} century
in the USA. Serendipitously, this also created a systemic limit on the ratio of money to bank
originated lending within the monetary system.  
While the sale of Asset Backed Securities does not allow any individual bank to 
exceed this ratio at any given point in time, as the process evolves the banking system itself
exceeds it as loans are moved outside the constraints provided by regulatory capital
or reserve regulation, thereby creating a mechanism for unconstrained 
growth in commercial bank originated lending. 
\subsection{The infiltration of Debt Instruments into the Capital Reserve}
While the asset backed security problem explains the dramatic growth in
banking sector debt that has occurred over the last
three decades, it does not 
explain the accompanying growth in the money supply. Somewhat uniquely 
of the many regulatory challenges that the banking system has created down the
centuries, the asset backed security problem, in and of itself
does not cause the banks, or the banking system to "print money".
\par
The question of what exactly constitutes money in modern banking systems is a
non-trivial one. As the examples above show, bank loans create money in the form
of bank deposits, and bank deposits can be used directly for monetary purposes
either through cheques or more usually now direct electronic transfer. For economic
purposes then, bank deposits can be regarded as directly equivalent to 
physical money.
The reality within the banking system however is somewhat more complex, in that transfers between bank
deposits must be performed using deposits in the clearing mechanisms, either
through the reserves at the central bank, or the bank's own asset deposits at
other banks. Nominally limits on the total quantity of central bank reserves should
in turn limit the growth in bank deposits from bank lending, but it is clear from the monetary 
statistics that this is not the case.  
\par
Individually commercial banks are limited in the amount of money they can 
lend. They are limited by any reserve requirements for their deposits,
by the accounting framework that surrounds the precise
classification of assets and liabilities within their locale,
and by the ratio of their equity or regulatory capital to their outstanding,
risk weighted loans as recommended by the Basel Accords. However none of these limits
is sufficient to prevent uncontrolled expansion.
\par
Reserve requirements at the central
bank can only effectively limit bank deposits if they apply to all accounts in the system,
and the central bank maintains control over any mechanisms that allow individual
banks to increase their reserve holdings. This appears not to be the case. In the US system for example, 
only Net Transaction Accounts(non-interest
chequing accounts) carry a reserve requirement, there are no restrictions on inter-bank lending
of reserves, and banks can increase their reserve holdings by depositing treasury instruments
which can be bought on the open market.
\par
Basel capital restrictions can also limit bank lending.
Assets(loans) held by banks are classified by type, and have accordingly different percentage
capital requirements. Regulatory capital requirements are divided into two 
tiers of capital with different provisions and risk categorisation applying
to instruments held in them.
To be adequately capitalised under the Basel accords, a bank must maintain a 
ratio of at least 8\% between its Tier 1 and Tier 2 capital reserves, and 
its loans. Equity capital reserves are provided by a bank's owners 
and shareholders when the bank is created, and exist to provide a buffer
protecting the bank's depositors against loan defaults. 
\par
Under Basel regulation, regulatory capital can be held in a variety of instruments, depending
on Tier 1 or Tier 2 status. It appears that some of those instruments,
in particular subordinated debt and hybrid debt capital instruments, can represent debt
issued from within the commercial banking system.
\begin{quote}
\begin{description}
\item[] Annex A - Basel Capital Accords, Capital Elements
\begin{description}
\item[] Tier 1
\begin{description}
\item[(a)] Paid-up share capital/common stock
\item[(b)] Disclosed reserves
\end{description}
\end{description}
\begin{description}
\item[] Tier 2
\begin{description}
\item[(a)] Undisclosed reserves
\item[(b)] Asset revaluation reserves
\item[(c)] General provisions/general loan-loss reserves
\item[(d)] Hybrid (debt/equity) capital instruments
\item[(e)] Subordinated debt
\end{description}
\end{description}
\end{description}
\end{quote}
\par
Subordinated debt is defined in Annex A of the Basel treaty as:
\begin{quote}
(e) Subordinated term debt: includes conventional unsecured subordinated debt capital 
instruments with a minimum original fixed term to maturity of over five years and 
limited life redeemable preference shares. During the last five years to maturity, 
a cumulative discount (or amortisation) factor of 20\% per year will be applied to 
reflect the diminishing value of these instruments as a continuing source of strength. 
Unlike instruments included in item (d), these instruments are not normally available to 
participate in the losses of a bank which continues trading. For this reason these 
instruments will be limited to a maximum of 50\% of tier 1. 
\footnote{Basel Capital Accords. International Convergence of Capital
Measurement and Capital Standards. (\url{http://www.bis.org/publ/bcbs107.pdf})}
\end{quote}
This is debt issued by the bank, in various forms, but of guaranteed long duration,
and controlled repayment. In effect, it allows Banks to hold borrowed money in 
regulatory capital. (It is subordinate to the claims of depositors in the event of Bank failure.)
The inclusion of subordinated debt in Tier 2 allows 
financial instruments created from lending to become
part of the regulatory control on further lending, creating a classic 
feedback loop.
This can also occur as a second order effect if any form of regulatory capital 
can be purchased
with money borrowed from within the banking system.\footnote{In 2007 Glitnir Bank in 
Iceland provided loans to shareholders in a subsidiary Byr Savings Bank to increase
its regulatory capital. Ten of the loans were subsequently ruled illegal as they
were made to children between the ages of 1 and 17. There does not appear however to have
been any other breach in banking regulations in connection with the loans.
\url{http://icelandreview.com/icelandreview/daily_news/?cat_id=16567&ew_0_a_id=351421}}
\par
The problem this creates arises from interactions with asset backed securities,
since it follows from the preceding argument that if the total quantity
of loans in the system was in some way fixed, it would not be possible to increase 
the other side of the balance sheet, regulatory capital and deposits.
Unlike the previous example though, the feedback loop through regulatory capital
does have money supply implications. We proceed as before to show
the flows between two banks.
\paragraph{Initial Conditions}
Bank A has made loans of \$1000 and Bank B has no loans outstanding.
For clarity, it is assumed that Bank A has no other loans except for the loan it is securitizing,
and that reserve requirements of 10\% apply to all accounts. 
As the Bank is making loans for mortgages, the regulatory capital risk weighting for
its loans is 50\%. For the initial loan book of \$1000, this requires a regulatory capital
holding of 10\% of  $1000 * 50\% = \$50 $.        
\par
Once again, this is the strictest set of conditions that can currently apply. 
\begin{table}[h]
\centering
{
\caption{Failure of Regulatory Capital Control}
%\caption{}
\label{table_2a}
\begin{tabular}{|lcc|ccc||cc|}
\hline
\multicolumn{8}{|c|}{Initial Conditions} \\
\hline
     &\multicolumn{2}{c|}{Assets}                      &\multicolumn{3}{c||}{Liabilities}&     &            \\
Bank & Loans         &  CB Reserve& Deposits& Income & Capital   & $\sum$ Deposits & $\sum$ Loans\\
\hline
A:   & 1000          &   50       &  1000   &        &  50       &  2000           & 1000        \\
B:   &               &  1050      &  1000   &        &  50       &                 &             \\
\hline
\end{tabular}
}
\end{table}
\par
\paragraph{Step 1:} 
Bank A creates an \$900 MBS, and sells it to the depositor at B
for \$990.\footnote{Public information on the exact proceeds of MBS sales 
does not appear to be available. All that is necessary for this process to operate
is that it is possible to sell the MBS for more than the face value of the
loans issued, whether that is achieved through loan fees, securitization 
commissions, or by some other
mechanism.  The amount of profit generated from the instrument, will 
determine the time period of the subsequent money supply expansion.}
The \$90 profit is treated as a liability to share holders and held in a income account.
\begin{table}[h]
\centering
{
\caption{Failure of Regulatory Capital Control}
\label{table_2b}
\begin{tabular}{|lcc|ccc||cc|}
\hline
\multicolumn{8}{|c|}{Step 1} \\
\hline
     &\multicolumn{2}{c|}{Assets}                      &\multicolumn{3}{c||}{Liabilities}&     &            \\
Bank & Loans         &  CB Reserve& Deposits& Income & Capital & $\sum$ Deposits & $\sum$ Loans\\
\hline
A:   &  100          &     1040   & 1000   &    90   &     50    &  2000           & 1000        \\
B:   &               &       60   &   10   &         &     50    &                 &             \\
\hline
\multicolumn{8}{|l|}{\emph{Mortgage Backed Securities: }} \\
\multicolumn{8}{|l|}{\emph{  \$900 owned by depositor at Bank B}} \\
\hline
\end{tabular}
}
\end{table}
%\newpage
\paragraph{Step 2:} 
By paying \$80 of the resulting fees from the MBS sale to its employees in salaries and bonuses, 
Bank A can increase its deposits to \$1080. 
It also borrows \$10 as a subordinated loan from the depositor at Bank B and places this loan in its Tier
2 regulatory capital holdings. 
It could equally have used the \$10 from profits to directly increase regulatory capital, and the same outcome
would occur.
\begin{table}[h]
\centering
{
\caption{Failure of Regulatory Capital Control}
\label{table_2c}
\begin{tabular}{|lcc|ccc||cc|}
\hline
\multicolumn{8}{|c|}{Step 2} \\
\hline
     &\multicolumn{2}{c|}{Assets}                      &\multicolumn{3}{c||}{Liabilities}&     &            \\
Bank & Loans         &  CB Reserve& Deposits& Income & Capital & $\sum$ Deposits & $\sum$ Loans\\
\hline
A:   &  100          &     1050   & 1080   &    10   &     60    &  2000           & 1000        \\
B:   &               &       50   &    0   &         &     50    &                 &             \\
\hline
\multicolumn{8}{|l|}{\emph{Mortgage Backed Securities: }} \\
\multicolumn{8}{|l|}{\emph{\$900 owned by depositor at Bank B}} \\
\hline
\end{tabular}
}
\end{table}
\paragraph{Step 3:} 
Bank A can now create a new loan of \$980 which is
deposited at Bank B. As a result the total amount of bank deposits in the system, and the 
corresponding amount of lending increases. Note that A is overcapitalised and over-reserved
for this loan, and the actual limit on the size of the loan is the total amount on deposit at Bank A.
However, once the loan is deposited at Bank B, Bank A is under-reserved. This shortfall however can
be covered from the interbank lending market in the short term, and by increasing its deposit
holdings at the central bank in the longer term by purchasing and depositing US treasuries.
It could also be resolved by re-classifying a portion of its deposit accounts so that they
were no longer classified as net transaction accounts and thus no longer incurred a reserve requirement.
\begin{table}[h]
\centering
{
\caption{Failure of Regulatory Capital Control}
\label{table_3c}
\begin{tabular}{|lcc|ccc||cc|}
\hline
\multicolumn{8}{|c|}{Step 3} \\
\hline
     &\multicolumn{2}{c|}{Assets}                      &\multicolumn{3}{c||}{Liabilities}&     &            \\
Bank & Loans         &  CB Reserve& Deposits& Income & Capital & $\sum$ Deposits & $\sum$ Loans\\
\hline
A:   &  1080         &       70   & 1080   &    10   &     60    &  2060           & 1980        \\
B:   &               &     1030   &  980   &         &     50    &                 &             \\
\hline
\multicolumn{8}{|l|}{\emph{Mortgage Backed Securities: }} \\
\multicolumn{8}{|l|}{\emph{\$900 owned by depositor at Bank B}} \\
\hline
\end{tabular}
}
\end{table}
\begin{center}
\emph{As a result of the subordinated loan, in combination with the movement of money around the system
with the Mortgage Backed Security, bank deposits are created in
the banking system, outside of central bank regulation.}
\end{center}
\par
Over the eight years that figures are available for regulatory capital holdings 
for American banks for example\footnote{Data taken from FDIC Call
Reports for American Banks available at
\url{https://cdr.ffiec.gov/public/PWS/DownloadBulkData.aspx}},
the total amount of regulatory capital has approximately
doubled from \$629,169,018 in 2001 to \$1,191,116,687 in 2008.
\par
Although the total money supply (deposits) in the example is increasing,
the amount on deposit at Bank B is slowly shrinking. This would not occur if Bank A
sold the subordinated debt to a depositor at its own Bank, and would be masked if
Bank B was also engaged in subordinated debt manipulations. This demonstrates another
systemic problem in that institutions
that do not engage in aggressive lending practices are 
at a competitive disadvantage to those that do, creating the very real danger that over time 
bad banks will drive out the good.  Kara\cite{kara.2011} provides some evidence that this
is indeed the case by examining pricing differences between banks that are more actively engaged
in the securitization markets.
\par
Foundational economic textbooks such as Mankiw\cite{mankiw.1997}, are very specific about
money creation, stating that ``the Federal Reserve is responsible for 
controlling the supply of money in the economy''. Unfortunately, as this 
example demonstrates, this theoretical control is dependent upon assumptions 
about implementation details of the current
reserve banking system that do not appear to be correct. Central Bank reserve requirements in
particular could be used far more effectively than they currently are, and
the presence of any non-reserved accounts within a banking system presents a curious oversight 
in this context.
\par
Another observation that can be made is
that excessive profits within the commercial banking system should always be regarded
with some suspicion, especially if aggregate statistics also show that the money and loan supplies are
"endogenously increasing". Any malfunction within the regulatory framework
will preferentially favour the institutions performing it above those who are not by
providing an excess of profits, and result in supra-normal growth for those institutions.
This is likely to occur whether or not the individuals controlling the
banks are aware of the precise nature of the malfunction, purely from the role of money as an economic signal.
This should not be taken as an argument per se for 
direct control of bank profits, since that would be merely treating the symptoms;
but it is an observation that the precise source of profits from all banking
activity deserves careful scrutiny.           
\subsection{Interbank Lending}
Interbank lending is a necessary part of the banking system and is required in order to cover
the accounting imbalances created by normal short term flows of deposits between banks, 
and had already 
emerged as a feature of the pre-banking goldsmith era in Europe. However, it presents regulatory
challenges of its own, as can be inferred from its effective appearance on both sides of the equation in
Figure \ref{fig_basel}. A loan to a bank is represented as
a liability, and counted with deposits and capital; the money received from that loan is classed
as an asset, and is included in the bank's own cash holdings with their loans and other assets.
A race condition appears to exist 
if while the loan is still present on the bank's account, the bank is able to disperse funds to
customer deposits (say by paying salaries and bonuses), allowing additional
bank deposit expansion.
\par
When occurring as part of day to day operations, this may represent
a very small money creating leak. However the growth of interbank lending in some banking systems
suggests a more systemic problem may be occurring. 
\section{Empirical Analysis}
\subsection{United States of America}
\begin{figure}[h]
\begin{center}
\includegraphics[width=10cm]{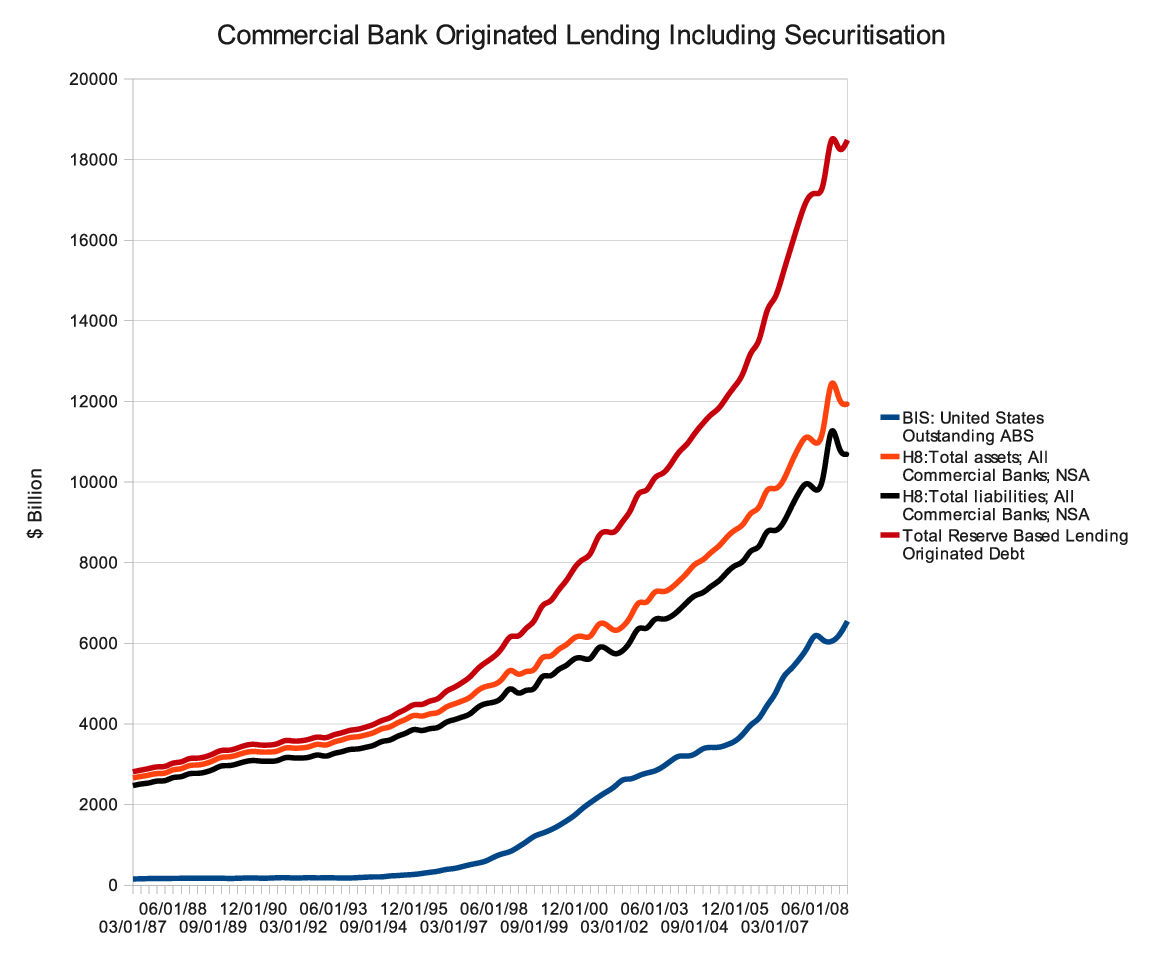}
\vspace*{8pt}
\caption{Estimate of USA Commercial Banks total originated debt, vs deposits}
\label{fig_absusassets}
\end{center}
\end{figure}

Continuous growth in the deposits and loans
controlled by the American commercial banks can be seen in the 
Federal Reserve statistical tables shown in Figure \ref{fig_absusassets}.
The total amount of outstanding loans originated by the commercial banks however
exceeds the quantity
of loans currently under their direct control, as it does not include
asset backed securities that have been resold.  Detailed figures on ABS issuance and ownership 
do not appear to be publicly available, but gross figures are available
from data provided by the Bank of International Settlements\footnote{\url{http://www.bis.org/statistics/secstats.htm}}.
\par
Figure \ref{fig_absusassets} also shows the aggregate of US outstanding 
ABS issuance, and US Banks assets. 
Figures on Mortgage Backed Security holdings by the US banks, which are the largest part of 
ABS issuances,
are newly available (since July 2009) in the Federal Reserve H.8 Table,
and show total MBS positions of \$977 billion in October 2009. 
This indicates that a ballpark estimate for the total excess issuance of 
loans by the 
commercial banks within the USA by 2009 was approximately \$5 trillion, or around 
half their total deposit liability, assuming that BIS figures include
securitized lending by the GSE's.  Pozsar\cite{pozsar.2010} however, in an extensive review
of commercial bank shadow banking activities includes a figure for total
liabilities in excess of bank lending of \$16 trillion which is hopefully an over estimate. 
Figure \ref{fig_absusassets} also
illustrates how the growth in total debt is increasing faster than
the total of underlying liabilities (money), as a direct consequence of 
loan securitization.\footnote{It is not known exactly where the ABS and MBS instruments purchased
by the Federal Reserve as part of the Troubled Assets Relief Program(TARP) feature in these
statistics.}
\subsection{Iceland}
Iceland is of interest as an example of a very small banking system,
operated using Basel rules, with an independent currency. 
Historically Iceland has experienced continuous stability problems with its banking system
since before the break up of the Bretton Woods agreement. 
Exceptionally high rates of monetary growth followed the breakup, and
consumer price inflation during the late 70's and early 80's in Iceland exceeded 50\%
for several years, peaking at just under 85\% in 1983\cite{andersen.1998}. 
A number of measures were introduced as a response, including
in 1979 the introduction of index linked mortgages, which indexed principle repayments to the CPI. In 1983
a nationally negotiated agreement imposed
a temporary suspension of wage indexation, and wage ceilings. 
Following this inflation subsided to annual rates during the 1990's that ranged 
between 5-10\%\footnote{The Annual Report of the Icelandic Central Bank(Sedlabanki Islands) can be found
online from 1997 at \url{http://www.sedlabanki.is/?PageID=178}, and from 1961 at the National
Archive and Central Bank Library}.
\par
\begin{figure}[h]
\begin{center}
\includegraphics[width=7cm]{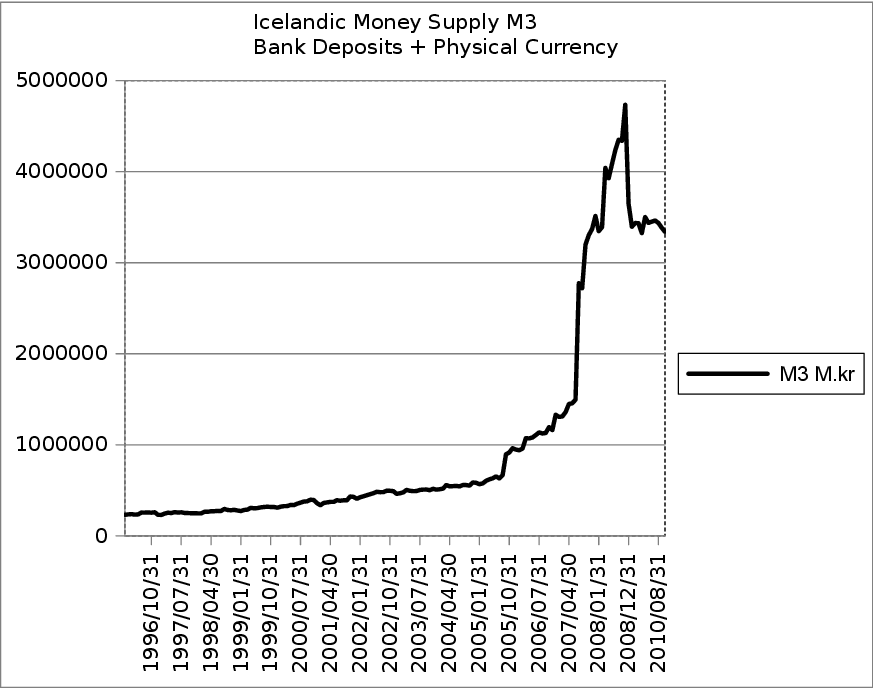}
\vspace*{8pt}
\caption{Icelandic Money Supply 1996 - 2010}
\label{fig_iceland}
\end{center}
\end{figure}
The Icelandic banks were deregulated and privatised at the end of the 1990's.
Figure \ref{fig_iceland}
shows the behaviour of the M3 money supply measure for Iceland since
1996. \footnote{The Icelandic M3 measure is the sum of all bank deposits and physical currency.}
Following deregulation an accelerated growth in money on deposit with the banks 
culminated in 2007 during which year
the money supply doubled. The proximate cause for the growth as discussed in the
Special Report to the Icelandic Parliament\cite{sic.2010} was an extraordinary
growth in lending by the Banks, accompanied by an expansion in the money supply of
ten times in a 7 year period as shown in Figure \ref{fig_iceland}.
\par
Whatever economic consequences might be expected to follow from bank privatisation,
a ten times increase in the total supply of money 
does not accord with either standard economic theory,
or the presumption of central bank control over the money supply.
In attempting to deal with the situation
the Icelandic central bank followed textbook
recommendations, and raised interest rates to over 15\% in order to stop
the monetary expansion.
This unfortunately had the reverse effect to that intended since it acted to
attract foreign deposits to the system, allowing the Icelandic banks
to further increase the amount of deposits that could be used to support
the loan supply, and allowed a small number of favourably
connected Icelandic businesses to engage in an international borrowing and investment spree,
part of which was used to increase regulatory capital holdings in the three main
Icelandic Banks.  
The regulatory capital growth was accompanied by substantial
inter-bank lending, and lending to business associates. 
\par
All three of the regulatory problems outlined
here appear during the bubble. Then Kaupthing (now Arion Banki) held a rapidly increasing
subordinate loan in their capital reserve from 1998, all three banks encouraged employees
to buy share capital, and made loans available for this purpose(illegally), and interbank lending
between the banks increased considerably\cite{sic.2010}.
\par
It is clear from the data of the period that there were no effective limits on 
lending within the Icelandic regulatory framework, but it is worth exploring 
what would have happened if there had been. The amount of business loans made by the Icelandic
banks would have been constrained by the limit. They may still have been made
imprudently, and consequent loan default would have been problematic, but there would
not have been the accompanying quantitative monetary expansion, and housing speculation triggered
by the rapid increases in property prices that accompanied the bubble. Post collapse, the domestic situation
in Iceland resembles Japan in that the majority of households are in  negative equity, and hold 40 year negative
amortisation index-linked mortgages.
\section{Credit Bubbles Revisited}
At the macro-economic level of the gross statistics of money and loan 
supply to the economy, the reserve banking system creates a complex interplay between 
money, debt, supply and demand for goods, and the general price 
level. Rather than being constant, as implied by theoretical descriptions,
money and loan supplies are constantly changing at a rate dependent on 
the average loan period, and a complex of details buried in the implementation 
and regulation of any given banking system. 
\par
Since the majority of loans are made for years at a time,
the results of these interactions play out over a long enough time scale
that gross monetary features of regulatory failure, such as continuous asset price
inflation, have come 
to be regarded as normal, e.g. "House prices always go up".
The price level however is not only dependent on purely monetary factors, but
also on the supply and demand for goods and services\footnote{Including financial assets
such as shares.}, which requires that estimates of the real price
level versus production be used as described by Dwyer\cite{dwyer.1999}. As a 
simplification, if constant demand for 
goods and services is assumed as shown in Table \ref{Table:msvsgs}, 
then there are 
two possible causes of price inflation, either the money supply available
to purchase the good in question has increased, or the supply of the good
has been reduced.\footnote{Population increases are also a factor, but are usually relatively
slow compared to the other factors. Understandably, underlying changes to the money
supply itself are an unwanted complication to many theoretical economic analyses, but 
Bordo \cite{bordo.2003} provides a more
nuanced analysis of price behaviour during the gold standard period supporting
the simplistic relationship shown. The claim by Fisher\cite{fisher.1911} that the velocity of circulation of money
can affect the price level, dates from before the multiplicative effect of loan re-deposit was fully 
understood, and is incorrect.} Critically, the former is simply a mathematical effect, 
whilst the latter is a useful signal, providing economic information on relative supply and 
demand levels that can be
used locally by consumers and producers to adapt their behaviour.
Purely arbitrary changes in both
the money and the loan supply that are induced by 
the mechanical operation of the banking system fail to provide any economic benefit, and by
distorting the actual supply and demand signal can be actively harmful.

\begin{table}[h]
\centering
\begin{tabular}{lll}
Money Supply &   Product Supply  & Price level  \\
\hline
Increases    &   Constant        &   Inflation \\
Decreases    &   Constant        &   Deflation \\
Constant     &   Increases       &   Deflation \\
Constant     &   Decreases       &   Inflation \\
\end{tabular}
\caption{Money and Productivity effects on Price Level}
\label{Table:msvsgs}
\end{table}
\par
Credit bubbles are often explained as a phenomena of 
irrational demand, and crowd behaviour. However, this explanation
ignores the question of why they aren't throttled by limits on the loan supply?
An alternate explanation which can be offered is that 
their root cause is periodic failures in the regulation of the loan and money supply within 
the commercial
banking system. In the current case the introduction of widespread securitized lending
allowed a rapid increase in the total amount of lending available from the banking system
and an accompanying if somewhat smaller growth in the money supply.
Channeled predominantly into property lending, the increased availability of money from 
lending sources, acted to increase house prices creating rational speculation on their increase,
and over time a sizeable disruption in the market pricing mechanisms for all goods and services
purchased through loans. Monetary statistics of this effect such as the CPI for example, were
however at least 
partially masked by production deflation from the sizeable productivity increases over the last decades.
Absent any limit on 
the total amount of credit being supplied, the only practical limit on 
borrowing is 
the availability of borrowers and their ability to sustain 
the capital and interest repayments demanded for their loans. 
\par
Owing to the asymmetric nature of long term debt flows there is a tendency for money
to become concentrated in the lending centres, which then causes liquidity
problems for the rest of the economy. Eventually repayment problems surface,
especially if the practice of further borrowing to repay existing loans is allowed,
since the underlying mathematical process is exponential.
As general insolvency as well as a consequent Fisher\cite{fisher.1933} 
debt deflation occurs,
the money and loan supply contracts as the banking system removes loan capacity
from the economy either from loan repayment, or as a result of bank failure. 
This leads to a domino effect as businesses that have become dependent on continuously 
rolling over debt fail and trigger further defaults. Monetary expansion and 
further lending is also constrained by the absence of qualified borrowers, and 
by the general unwillingness to either lend or borrow that results from the 
ensuing economic collapse. Further complications, as described by Bernanke\cite{bernanke.1990},
can occur when interactions between
currencies are considered, in particular in conjunction with gold based capital
regulation, because of the difficulties in establishing the correct ratio of
gold for each individual currency and maintaining it, in a system where lending 
and the associated money supply are continually fluctuating and gold is also
being used at a national level for international debt repayments.
\par
The debt to money imbalance created by the widespread, and global, sale of 
Asset Backed
securities may be unique to this particular crisis. Precise details are hard 
to obtain, but the Bank of International Securities provides national 
aggregates that suggest the world wide total of outstanding securitized debt
to be approximately \$25 trillion US dollars for the period for which statistics
are available.  Within the Euro zone there also appear to have been
considerable variations in the local regulation of lending and loan sales, and this is presumably
creating some degree of internal stress within the euro based economies, although a much larger
issue there is the different deposit expansion rates within their individual banking 
systems\footnote{OECD data for 1999-2009 indicates a range of expansions between 1.3 and 3.0 within
the Eurozone's constituent banking systems. (Eurozone OECD Bank Profitability Statistics, Income Statement and Balance Sheet (Source: OECD.StatsExtracts))}.
\par
Although asset backed security issuance dropped considerably in 2008,
as the resale markets were temporarily frozen,
current stated policy in several countries, including the USA and the United Kingdom,
is to encourage further securitization to assist the recovery of the banking
sector. Unfortunately this appears to be succeeding. 
\section{Conclusion}
The Banking System occupies a unique place in the economy acting as it does
on the supply of both money and credit, and as has been seen repeatedly
with a unique ability to disrupt it. 
Unfortunately a complete understanding of its behaviour does
not appear to have ever been correctly formulated by economic theorists.
The Basel framework resting on principles of risk assessment, and capital protection
does not appear to have been any more successful at providing systemic stability than any of its predecessors.
A complete and verifiable model of the banking system under the variety of regulatory
frameworks in use today should be regarded as a research priority, and this is where a complex systems
approach can make a valuable contribution, by providing the theoretical basis for a 
complete evaluation of its systemic
behaviour and models of its interaction with the price level and consequent feedback effects.
\par
Such a model could be particularly helpful in guiding corrective interventions.
Recommendations based on Keynesian theory, and in particular on the experience of the
1930's Great Depression assume that the underlying banking system and the problems
being experienced with it are the same as were seen during that period. Today's banking 
system is however structured very differently, and the regulatory framework has changed
sufficiently that direct comparisons between the two systems are unlikely to be substantive.
In particular there is no evidence to suggest that the debt/money imbalance introduced by Asset Backed Securities was 
present at that time.  In today's banking system attempting to correct bank stability
problems by injecting more money into the economy may well backfire, as they can cause proportionally 
more debt to be created, further increasing leverage within the economy.
\par
Finding effective solutions to the problems of the current banking system presents considerable challenges,
and putative reformers should bear in mind that this is by any definition an economically critical
system, where seemingly insignificant changes can have extremely far reaching effects as the introduction
of securitization amply demonstrates. A far more rigorous approach to evaluating proposed
changes is urgently needed, and drawing on experience from the safety critical and fault tolerant systems
community may be useful in that regard.  
\par
Banning asset backed securities at this point in time would effectively
remove over \$1 trillion of annual credit issuance, causing a new global credit crisis,
with knock on effects on other forms of borrowing. 
In the longer term, since asset backed securities are
money supply neutral once they are outside of the banking system, and since
debt has an intrinsic decay function over time as it is repaid, their effect on the
economy will slowly decline so the possibility of slowly removing them from the
system exists.  Similar considerations apply to stabilising the money supply
by properly regulating regulatory capital and reserve requirements, in the absence
of action on asset backed securities this would only serve to accelerate the growth in the money to debt 
imbalance in the system. Only a system level intervention to tackle both problems 
simultaneously would be likely to succeed in the long term. 
\par
The changes introduced into the Banking system by the Basel treaties,
appear to have been intended to stabilise the banking system against previous problems, 
by shifting regulatory emphasis to default risk and providing capital buffers as
protection. However these changes concentrated on the risk of default at individual banks,
there appears to have been relatively little attention given, either to the behaviour of the system as 
a whole, or the risk of its mechanisms being subverted. Unfortunately, due to the oversight with 
respect to the interaction of
regulatory capital debt instruments,  and the increasingly widespread use of Asset Backed Securities,
the eventual result may have been to protect the individual members of the system, 
the banks, at the expense of the structural integrity of the system itself. 
\section{Acknowledgements}
The author would like to thank the many people who have assisted with this paper. In particular
Dr. Lorin Wilde, Dr. Brendan Sheehan, and the journal reviewers for their extremely valuable feedback 
and criticism which has considerably improved it, 
Ilya Kuntsevich and Dr. Iman van Lelyveld for their assistance with understanding the Basel Accords, 
Dr. J\'on Thor Sturluson for his assistance on the relationship between liquidity and fractional 
reserve operations under the Basel regulatory framework, Dr. Marco Raberto for suggesting a 
double entry bookkeeping presentation, and Valborg Stef\'ansd\'ottir and Anton Holt at the Library of the 
Central Bank of Iceland for their kind assistance with access to information on the Icelandic Banking System. 
It should not be presumed that they necessarily 
agree with my conclusions, and the author takes sole responsibility for any remaining errors.
\bibliography{limits}
\raggedright
\bibliographystyle{unsrt}

\end{document}